\documentclass[12pt]{article}

\usepackage{amsmath}
\usepackage{amssymb}
\usepackage{graphicx}
\usepackage{color} 

\usepackage{fullpage}
\usepackage{url}

\usepackage{authblk}

\date{}

\begin{document}

\begin{flushleft}
{\Large
\textbf{Corruption Drives the Emergence of Civil Society}}

\vspace{0.3cm}
Sherief Abdallah$^{1,2,3,\ast}$, 
Rasha Sayed$^{1}$, 
Iyad Rahwan$^{4,2}$,
Brad LeVeck$^{5}$,
Manuel Cebrian$^{6,7}$,
Alex Rutherford$^{4}$,
James H. Fowler$^{5}$
\\ \vspace{0.3cm}
\bf{1} The British University in Dubai, Dubai, UAE
\\
\bf{2} School of Informatics, University of Edinburgh, Edinburgh, UK
\\
\bf{3} Cairo University, Cairo, Egypt
\\
\bf{4} Masdar Institute of Science \& Technology, Abu Dhabi, UAE
\\
\bf{5} University of California, San Diego, La Jolla, CA, USA
\\
\bf{6} NICTA, Victoria, Australia
\\
\bf{7} University of Melbourne, Victoria, Australia
\\
$\ast$ To whom correspondence should be addressed; E-mail: \texttt{shario@ieee.org}
\end{flushleft}



\begin{bf}
Peer punishment of free-riders (defectors) is a key mechanism for promoting cooperation in society \cite{boyd1992punishment,fehr1999cooperation,egas2008economics,axelrod2006evolution,ohtsuki2006simple}. However, it is highly unstable since some cooperators may contribute to a common project but refuse to punish defectors.
%
Centralized sanctioning institutions (for example, tax-funded police and criminal courts) can solve this problem by punishing both defectors and cooperators who refuse to punish.  These institutions have been shown to emerge naturally through social learning and then displace all other forms of punishment, including peer punishment  \cite{sigmund10nature,traulsen2012economic}.
%
However, this result provokes a number of questions.  If centralized sanctioning is so successful, then why do many highly  authoritarian states suffer from low levels of cooperation?  Why do states with high levels of public good provision tend to rely more on citizen-driven peer punishment?  And what happens if centralized institutions can be circumvented by individual acts of bribery?
%
Here, we consider how corruption influences the evolution of cooperation and punishment.  Our model shows that the effectiveness of centralized punishment in promoting cooperation breaks down when some actors in the model are allowed to bribe centralized authorities.
%
Counterintuitively, increasing the sanctioning power of the central institution makes things even worse, since this prevents peer punishers from playing a role in maintaining cooperation. As a result, a weaker centralized authority is actually more effective because it allows peer punishment to restore cooperation in the presence of corruption.
%
Our results provide an evolutionary rationale for why public goods provision rarely flourishes in polities that rely only on strong centralized institutions.  Instead, cooperation requires both decentralized and centralized enforcement.  These results help to explain why citizen participation is a fundamental necessity for policing the commons. \\
\end{bf}


A centuries-old debate exists on how to best govern society and promote cooperation: Is cooperation best maintained by a central authority~\cite{hobbes1960leviathan, hardin1968}, or is it better handled by more decentralized forms of governance~\cite{kropotkin1907mutual, dietz2003struggle}?  The debate is still unresolved, and identifying mechanisms that promote cooperation remains one of the most difficult challenges facing society and policy-makers today \cite{dietz2003struggle}. 

Decentralized, individual sanctioning of non-cooperators (also known as free-riders or defectors) is one of the main tools used by societies to promote and maintain cooperation \cite{nowak2006five}. Individuals can sanction free-riders implicitly via behavioral reciprocity (as in the case of the highly successful tit-for-tat strategy \cite{axelrod2006evolution}), or  explicitly via costly punishment \cite{fehr::2002}. Both of these forms of peer punishment have been widely studied using evolutionary models and behavioral experiments \cite{boyd1992punishment,fehr1999cooperation,egas2008economics,axelrod2006evolution,ohtsuki2006simple}.

Recently, however, Sigmund et al. \cite{sigmund10nature} showed that centralized institutions can have an evolutionary advantage over peer punishment because, unlike peer-punishers, these institutions may eliminate ``second-order'' free-riding. Second-order free-riders cooperate with other players, but they do not pay the cost of punishing defectors, and this can allow defectors to re-emerge \cite{panchanathan2004indirect,fowler2005human,dreber08nature}. To address this problem, Sigmund et al. present a model of ``pool'' punishment, where agents commit resources to a centralized authority that sanctions free-riders \cite{sigmund10nature,traulsen2012economic}. Pool punishment avoids the second-order free-rider problem because the centralized authority also punishes individuals who do do not punish, either directly or by contributing to the pool. This allows pool-punishers to quickly take over a population, displacing both free-riders and peer-punishers \cite{sigmund10nature}.  These advantages help to explain why human societies frequently delegate punishment to centralized institutions \cite{sigmund10nature,baldassarri11pnas,traulsen2012economic}. They also help explain why centralized institutions acquire an increasing monopoly over legitimate punishment over time by stigmatizing \cite{rosenbaum11nyt} and criminalizing \cite[p. 371-372]{hallam1821view} various forms of peer punishment.

However, the dominance of pool punishment in the Sigmund et al. model~\cite{sigmund10nature} also creates three puzzles. First, the results imply that increasing the severity of centralized pool-punishment always increases cooperation. Yet many authoritarian states, which have the ability to severely punish citizens, suffer from low levels of participation and public goods provision \cite{acemoglu2012nations}. Meanwhile states with high levels of public goods, such as western democracies \cite{acemoglu2012nations,deacon2009public,lake2001invisible}, typically limit the government's ability to punish individuals and tolerate more forms of peer punishment.

Second,  centralized pool punishment quickly takes over a population and completely displaces peer punishment \cite{sigmund10nature} in the  Sigmund et al. model~\cite{sigmund10nature}, but many (if not most) societies exhibit a mix of centralized and decentralized punishment strategies. Even in societies with centralized punishment, citizens engage in costly acts of protest against agents who harm the public good. As recent events---from the Occupy protests to the Arab Spring---illustrate, this occurs even when the government punishes protestors \cite{harcourt2011occupy,morsy13ahram, moghadam2012globalization}. What unmodeled factors might allow peer punishment to evolve alongside centralized enforcement institutions---even when these institutions are actively hostile towards various forms of peer punishment?

Third, the Sigmund et al. model~\cite{sigmund10nature} assumes that the centralized authority punishes all forms of peer punishment. This is because peer-punishers in their model, by definition, do not contribute to the centralized authority. However, many societies with centralized enforcement also recognize certain forms of peer punishment as legitimate. For instance, civil litigation, jury duty, anti-incumbent voting, and other forms of political participation are also instances of altruistic peer punishment \cite{fowler2007beyond,grechenig2010punishment,smirnov2010behavioral}. In these and other cases, citizens engage in a \emph{hybrid} peer-pool punishment strategy. These individuals pay taxes to a central authority, but also engage in selective acts of peer punishment that are individually costly, but not punished by a central power.  Given all the costs they bear, it is unclear how such hybrid strategies may evolve.

Here we show that allowing for \emph{corruption} in the model can help explain both why societies want to limit the severity of centralized punishment, and why peer punishment frequently evolves alongside centralized punishment institutions. We investigate the effect of corrupt players who can bribe a central authority to avoid punishment. The results show that when pool-punishers dominate a system, the central authority becomes a single point of failure, which is highly vulnerable to corruption. This gives an opportunity for individuals playing a hybrid peer-pool strategy to evolve because peer punishment becomes relatively more effective under these circumstances by helping to increase the overall level of cooperation.


In sum, given the possibility of corruption, Leviathans can promote cooperation, but only if they also allow individuals to take action against actors who harm the public good. Our model therefore provides an evolutionary rationale for why public goods provision and cooperation rarely flourish in polities with strong centralized punishment alone. Instead, cooperation rests on an authority that protects a fundamental aspect of civil society, citizen participation in policing the commons \cite{putnam1994making, verba:2005vi}.

Our baseline model is a public good game (PGG) with both peer and pool punishment \cite{sigmund10nature}. The PGG is a simple model for studying contributions to a project with non-excludable positive externalities, which may include everything from the provision of social insurance to the protection of the environment. Let $M$ denote the population size and let $N\le M$ denote the number of individuals who are randomly chosen in a given round to play a public good game. In the game, each individual is faced with a choice: whether or not to contribute a fixed amount, $c>0$, to the common pool. Once each individual chooses her action, each individual will obtain $rc\frac{N_c}{N}$, where $r$ is a factor greater than 1, $N_c$ is the number of contributors to the common pool, and $N$ is the total number of participants (whether they contributed or not). If all individuals contribute, $N_c=N$, then the social welfare is maximized and each individual obtains $rc$. However, each actor gains an equal share of $rc\frac{N_c}{N}$, whether or not they contribute, making it a dominant strategy for each individual to free-ride by contributing 0 (the payoffs are written explicitly in the SI). 

The population includes $X$ \emph{cooperators}, who contribute $c$, and $Y$ \emph{defectors (free-riders)}, who do not. Consistent with previous work, we also assume the game is not compulsory and some players may choose not to participate in the PGG \cite{sigmund10nature, fowler2005altruistic, Hauert10052002, hauertwz2002replicator,semmann2003volunteering}. These \emph{loners} earn a fixed small pay-off, $\sigma$. In addition, $W$ \emph{peer-punishers} cooperate by contributing $c$ to the PGG, but also impose a fine, $\beta$, on each free-rider at a cost $\gamma$ \cite{nowak2006five}. In other words, each free-rider pays a total fine $\beta N_w$, where $N_w$ is the number of peer-punishers in the group, and every peer-punisher incurs an extra cost $\gamma N_f$, where $N_f$ is the number of free-riders in the group. We also have $V$ \emph{pool-punishers} who, instead of directly punishing free-riders, contribute a fixed amount, $G$, to a punishment pool before participating in the game and then contribute $c$ to the PGG. Those who do not contribute to the pool (both free-riders and peer-punishers) are then fined $BN_v$ each, where $N_v$ is the number of pool-punishers in the group.  We also introduce $C$ \emph{corruptors} to the model. A corruptor pays the central authority a fixed fee $KG$ to avoid being punished for not contributing to the PGG (this only makes sense if the fee is less than the total contributions paid by pool-punishers, $KG<G+c$). 

We study the equilibria of fully mixed populations of fixed size $M$ and variable composition by computing the pay-offs obtained by players using these strategies, assuming agents play in randomly sampled groups of size $N$. The difference in payoffs, together with the parameter $s \geq 0$, determine the rate at which individuals with lower payoffs are replaced by types with higher payoffs.  As in other evolutionary models, this process can be interpreted either as evolution or social learning. We also allow for random switching of strategies with a mutation rate $\mu \geq 0$. We derive equilibria as the long-run distribution of different strategies both analytically (in the limit of strong imitation) and using numerical simulation.  For details, see the Supplementary Information (SI).

Figure \ref{fig-runs} shows sample runs of a numerical simulation of the model. Without corruption, pool punishment eventually takes over the population \cite{sigmund10nature}, and does so even earlier when $B$, the severity of second order pool punishment, is higher.

The situation changes dramatically when we introduce corruptors. Pool punishment is no longer a dominant strategy, as shown in a sample simulation run (Figure \ref{fig-runs}(c)). Interestingly, Figure \ref{fig-runs}(d) shows that weakening pool punishment (lowering the fine $B$) allows peer-punishers to re-emerge as a relatively stable strategy that restores cooperation in the presence of corruption.

\begin{figure*}
\centering
\includegraphics[width=6in]{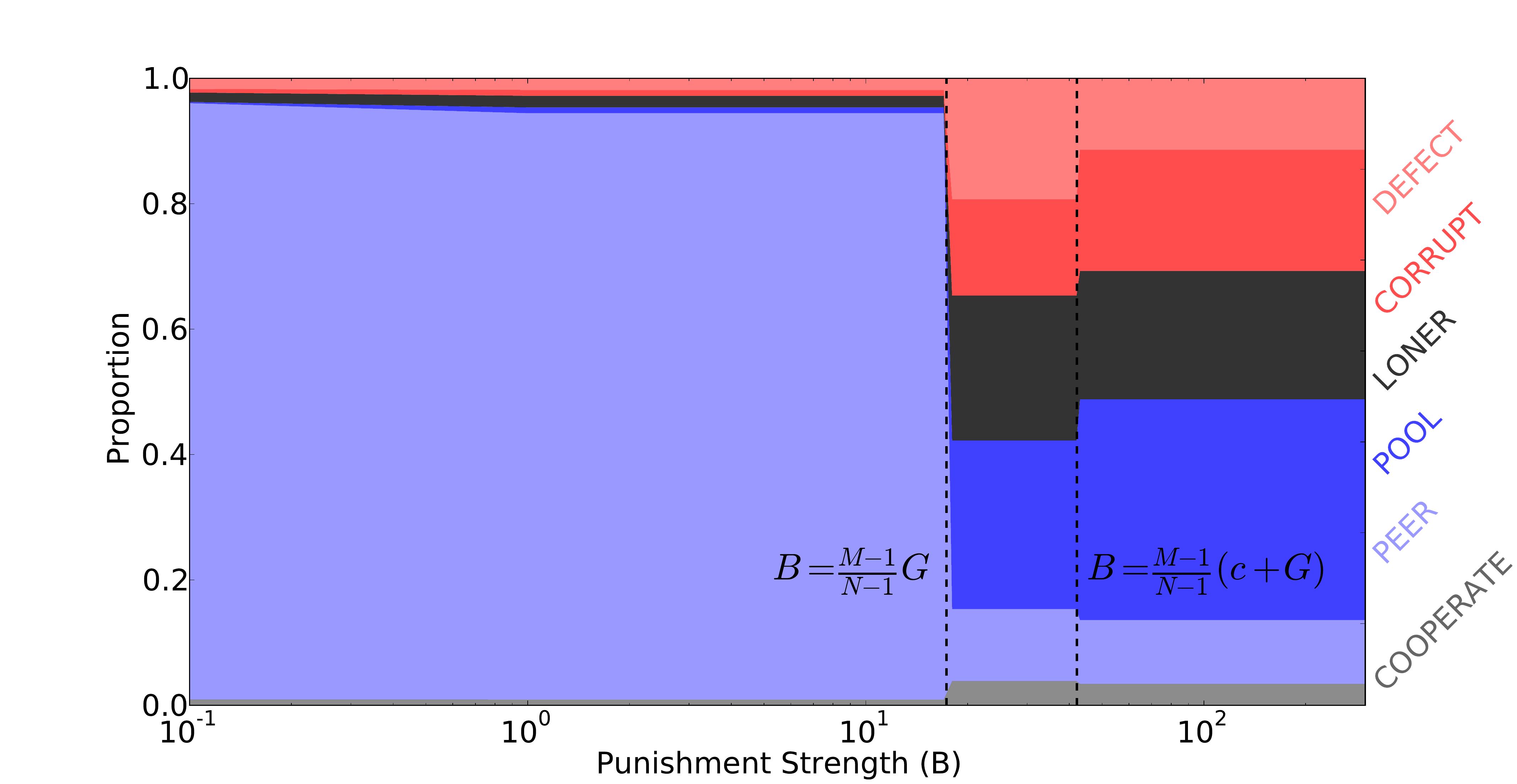}
\caption{Stationary distributions of strategies as a function of institutional punishment severity (parameter $B$), with the corruptor strategy included in the set of available strategies. Here we observe the adverse effect of institutional punishment. The greater $B$, the greater the percentage of corruption. A clear phase transition happens when $B>\frac{M-1}{N-1}G$, when the expected punishment exerted by a single pool-punisher (in a sample of N) exceeds the punishment cost for the pool-punisher, $G$.  This allows a pool-punisher to severely suppress peer-punishers, which in turn allows corruptors, defectors, and loners to grow in the population.} 
\label{fig-phase}
\end{figure*}

We investigate this further in Figure \ref{fig-phase}, which shows the proportion of different strategies as a function of second-order punishment severity. For low values of $B$, peer-punishers dominate and prevent the corruptor strategies from gaining ground. As $B$ increases, peer-punishers disappear and pool-punishers become more prevalent. However, with even higher values of $B$, the prevalence of corruptors also increases. This causes the total number of cooperative individuals to decline. We confirm these results by analytical computation of the long run frequencies of strategies in the $(X,Y,Z,V,W,C)$ sub-population (for methods see SI). With low $B$, the frequencies, respectively, are $\frac{1}{M+7}\left[1,2,2,1,M,1\right]$, confirming the clear dominance of peer-punishers (with a population of $M=100$ this is approximately $\left[ 0.01,0.02,0.02,0.01,0.93,0.01\right]$). Strong central punishment, however, yields the distribution $\left[0.034,0.114,0.204,0.352,0.102,0.193\right]$, i.e. ineffective (corrupted) pool-punishers dominate, followed by loners and corruptors.

Strong centralized punishment allows corruptors to exploit pool-punishers in two ways: Pool-punishers contribute to a public good, while at the same time funding a flawed institution that corruptors use to their advantage. Weak centralized punishment, on the other hand, provides an opportunity for peer-punishers to counteract both corruptors and defectors.

Lastly, we introduce $H$ \emph{hybrid punishers}. In addition to contributing $c$ to the public good, individuals using this strategy pay both $\gamma$ to punish defectors directly, as well as $G$ to the punishment-pool, and as such they are not punished by the central authority. Hybrid individuals can be thought of as upstanding citizens that pay their taxes, but also engage in forms of `legitimate' peer sanctioning. 

\begin{figure*}
\centering
\includegraphics[width=6in]{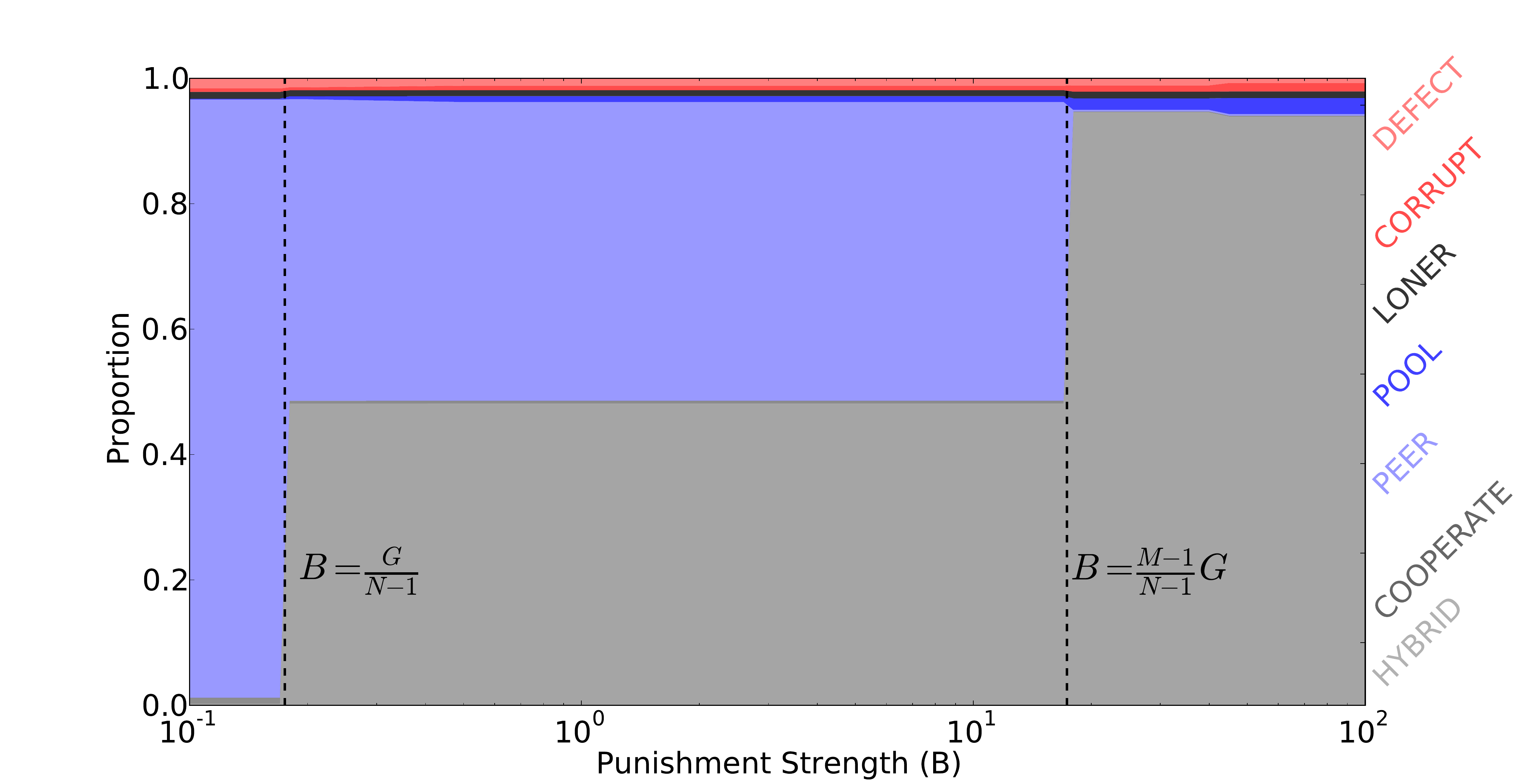}
\caption{Stationary distributions of strategies as a function of institutional punishment severity (parameter $B$), with both the corruptor and the hybrid strategies included in the pool of available strategies. Hybrid strategies contribute to the punishment pool, but also engage in peer punishment (without being sanctioned for it). As a result, increasing $B$ no longer back-fires, and the same level of cooperation is maintained. The hybrid strategy becomes dominant for $B>\frac{M-1}{N-1}G$, when the expected punishment exerted by a single peer-punisher (in a sample of N) exceeds the punishment cost for the pool-punisher, $G$. The first transition occurs for $B>\frac{G}{N-1}$, when the punishment imposed by a single pool-punisher exceeds the cost in a sample of $N$ players.}
\label{fig-peerpool}
\end{figure*}

Figure \ref{fig-peerpool} shows that, unlike peer punishment alone, this hybrid strategy dominates the population when centralized punishment is severe. This occurs even though the hybrid strategy pays a higher average cost compared to pool-punishers. Setting $M=100$, the long-run distribution of strategies in the $(X,Y,Z,V,W,C,H)$ sub-population is $$\left[ 0.001, 0.004, 0.008, 0.013, 0.004, 0.007, 0.96\right]$$ As a consequence, a high level of cooperation is maintained across all levels of centralized punishment.

Of course, one might wonder why individuals would create a second-order punishment institution in the first place, as Figure \ref{fig-peerpool} also shows that second-order punishment does not increase the overall level of cooperation; nor does it make cooperation significantly more stable than peer-punishment alone. Our relatively simple model is unlikely to fully answer this very general question, as we have left out many features which could cause centralized institutions to remain advantageous. For example, these institutions might aggregate views on who should be punished; and, this aggregation could cause perceptual errors (which are not in our model) to cancel out \cite{mclennan1998consequences}.

It is also possible that institutions may further evolve to deal with this remaining instability. Analytical results in the SI show that when second-order punishment is strong, hybrid punishers are only destabilized by neutral-drift towards pool-punishers (who then allow corruptors and defectors to emerge). Institutions may therefore want to screen and punish pure pool-punishers; and, it is interesting that many justice systems have evolved rules that fine people who merely pay their taxes, but do not register for various forms of hybrid punishment, such as jury duty.

Importantly, however, we have shown that simply adding the risk of corruption can help explain why centralized and decentralized forms of punishment frequently co-exist. No additional appeal to civic norms or civic culture is needed. Which is not to say these things do not exist, or that they do not further promote citizen participation in policing the commons. Rather, our model shows that independent of other virtues, peer-punishment strategies can have a fitness advantage over pool-punishment alone. In the face of corruption, peer and hybrid punishment strategies better promote cooperation because they are competitive. If one punisher fails to punish a corrupt individual, another might step in; and, this result may help to explain why polities who want to control corruption and promote cooperation often become more tolerant of various forms of decentralized sanctioning \cite{egorov2009resource}.

\section*{Acknowledgements}
M.C. is funded by the Australian Government as represented by the Department of Broadband, Communications and the Digital Economy and the Australian Research Council through the ICT Centre of Excellence program.

\section*{Author Contributions}
S.A., I.R., M.C. designed the research; S.A., R.S., A.R. implemented the simulations; S.A., R.S., I.R., M.C., J.H.F., B.L, A.R. analyzed the results and wrote the paper.

\section*{Competing financial interests}
The author(s) declare no competing financial interests.

\clearpage


\begin{figure*}
\centering
\includegraphics[width=5in]{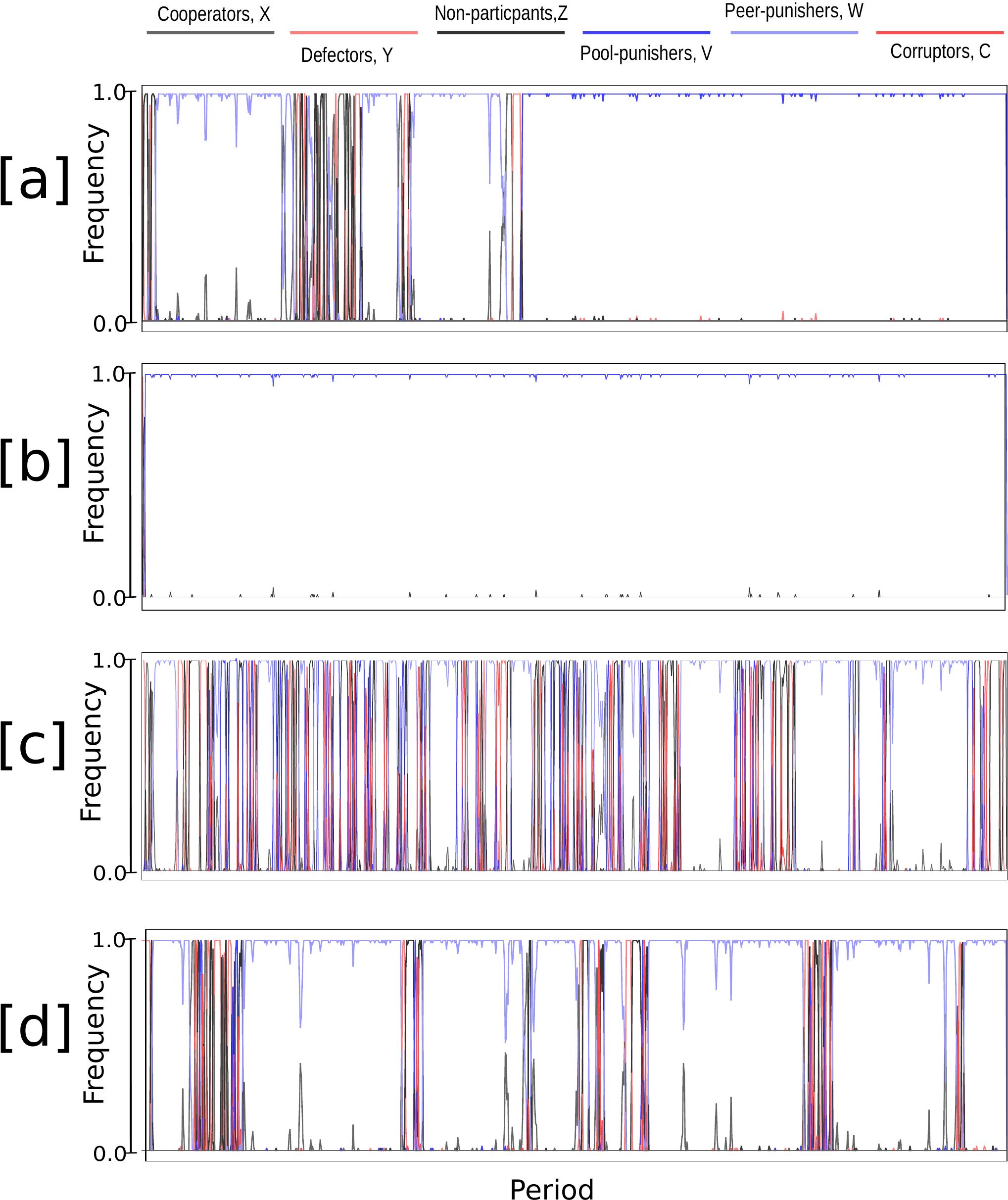}
\caption{Sample simulation runs showing the effect of the corruptor strategy. For all the runs, the following parameter values were used (please refer to the supplementary material for more details): 
$S=100000, M = 100, N=5, \mu= 0.001, \sigma= 1.0, C= 1.0, R= 3.0, \beta= 0.7, \gamma= 0.7,K=0.5$, and $G=0.7$. The severity of institutional punishment is controlled via parameter $B$ which is set to either $0.7$ or $7$. 
In (a), without the corruptor strategy, the results are consistent with the results reported in the previous work \cite{sigmund10nature}, where pool-punishers predominate. In (b), the predominance of the pool-punishers becomes decisive as the severity of the institutional punishment escalates. In (c), with the corruptor strategy added to the mix of available strategies, and with the severity of institutional punishment set to ($B=7$), the pool-punishers are no longer stable and cooperation deteriorates in general. Finally, in (d), as institutional punishment becomes more lenient, peer-punishers emerge and largely maintain cooperation ($B=0.7$), even in the presence of corruptors.}
\label{fig-runs}
\end{figure*}

\end{document}


\maketitle

\section{Calculation of Stationary Distribution}

Our model of cooperation follows the common formulation of evolutionary dynamics simulations~\cite{novakBook}. Specifically we consider a set of $M$ agents each subscribing to one of $d$ strategies. At each time step a random sample of $N$ agents are chosen to play a public goods game. The payoffs received by each agent are determined by the number of each type of strategy. At each time step 2 agents are randomly chosen and their payoffs are compared. The probability of one agent imitating the other is determined by a logistic function of the difference in payoffs and an imitation strength $s$. There is also a small probability $\mu$ that a randomly chosen agent will undergo a mutation to a different strategy.\\

In order to calculate the stationary distribution of strategies in our evolutionary dynamics we consider, in common with previous work on life-death processes~\cite{fudenberg}, the rates of transitions between homogeneous states in which all agents subscribe to a single strategy. Under deterministic dynamics these homogeneous states may be absorbing i.e. Once cooperation has collapsed and defectors have taken over, the system cannot return to a homogeneous state of cooperators. However random mutation allows mixing between homogeneous states via \textit{mutation} and subsequent \textit{fixation}.\\

Consider a population of agents each subscribing to strategy $X$. The probability that the system makes the transition to the state of all agents subscribing to a different strategy $Y$ depends on the product of two quantities;

\begin{enumerate}
	\item{The probability that a random mutation introduces an agent with strategy $Y$ ($\mu_{X,Y}$)}
	\item{The probability that this single mutant can invade the population and lead all agents to switch to strategy $Y$; this is known as the fixation probability ($\rho_{X,Y}$).}
\end{enumerate}

In this formulation we assume that the mutation rate is low so that each mutation event leads either to fixation of a new homogeneous state or reversion to the same homogeneous state before the next mutation event occurs. Therefore, at any given time, at most two strategies are present.\\

Addressing (1), mutations occur in the population at a rate $\mu$. The resultant strategy is chosen from the $d-1$ other strategies at random, giving a mutation probability

\begin{equation}
	\mu_{X,Y}=\frac{\mu}{(d-1)}
\end{equation}

Addressing (2), the fixation probability can be expressed explicitly from the product of the probability of each agent, after the first mutant agent, successively imitating the invading strategy. This requires a detailed description of the payoffs and imitation probabilities (section \ref{hard}). Alternatively, (2) can be inferred simply in the limit of strong imitation (section \ref{easy}).\\

Once we have an expression for the transition matrix between the homogeneous states, we can find the stationary distribution of the system of agents as the dominant eigenvector. This is a vector of values of size $d$ which represents the long run probabilities of finding the system in a given state. We require that the transition matrix $T$ be row normalised i.e. If the system is found in state $X$ it must either remain in state $X$ or transition to state $k\neq X$. Because the stationary distribution tells us the \textit{relative proportions} of each state and the fact that the mutation probability does not depend on the source or target states, the \textit{actual numerical value} of $\mu$ is not important and it is convenient to omit it from $T$.\\

For a simple system of $d=$ 3 states $X$,$Y$ and $Z$ representing cooperators, defectors and non-participants respectively, we can construct $T$

\begin{equation}
\renewcommand\arraystretch{2}
T = \begin{pmatrix}
	1-\frac{1}{2}\rho_{X,Y}-\frac{1}{2}\rho_{X,Z}&\frac{1}{2}\rho_{X,Y}&\frac{1}{2}\rho_{X,Z}\\
	\frac{1}{2}\rho_{Y,X}&1-\frac{1}{2}\rho_{Y,X}-\frac{1}{2}\rho_{Y,Z}&\frac{1}{2}\rho_{Y,Z}\\
	\frac{1}{2}\rho_{Z,X}&\frac{1}{2}\rho_{Z,Y}&1-\frac{1}{2}\rho_{Z,X}-\frac{1}{2}\rho_{Z,Y}\\
\end{pmatrix}
\end{equation}

The factor of $\frac{1}{2}$ corresponds to $\frac{1}{d-1}$. 
\subsection{Strong Imitation Limit}
\label{easy}
The individual entries of $T$ can be populated by simple arguments under the limit $s\rightarrow \infty$ (and under suitable conditions for other parameters such as punishment strength or cost) so that a strategy with a superior payoff will always be imitated and an inferior payoff will not. There are in fact only 3 possible values for the fixation probabilities $\rho_{i,j}$

\begin{enumerate}
	\item[$\boldsymbol{\rho_{i,j}=0}$: ] If $P_{j}<P_{i}$ for a single mutant with strategy $j$, then the mutation cannot invade and the fixation probability is 0.
	\item[$\boldsymbol{\rho_{i,j}=1}$: ] If $P_{j}>P_{i}$ for a single mutant with strategy $j$, then the mutation is beneficial and induces transition to a homegenous state $j$
	\item[$\boldsymbol{\rho_{i,j}=\frac{1}{2}}$: ] This is peculiar to a single cooperator attempting to invade non-participants. The non-participants receive a fixed payoff of $\sigma$ but a single cooperator will also receive a payoff $\sigma$ since she has no partner with which to participate in a PGG. At the next imitation event involving the mutant cooperator, the cooperator will have the opportunity to imitate a non-participant. Since the payoffs are identical, the cooperator will revert to a non-participant with probability $\frac{1}{2}$, but is equally likely to convert a non-participant to cooperation under a neutral drift. Once two or more cooperators are present, this strategy is dominant and they invade with probability 1.
\end{enumerate}

Our intuitive understanding of PGGs tells us that in the absence of punishment, free-riding always pays ($\rho_{X,Y}=1$) and that unilateral cooperation in the face of defection does not ($\rho_{Y,X}=0$). When cooperation is underway, it pays to participate ($\rho_{X,Z}=0$) and due to the argument above, cooperators are slow to take over non-participants ($\rho_{Z,X}=\frac{1}{2}$). Finally, if no-one is playing the PGG then something is better than nothing ($\rho_{Y,Z}=1$ and $\rho_{Z,Y}=0$). Therefore $T$ reduces to

\begin{equation}
\renewcommand\arraystretch{2}
{T=\begin{pmatrix}
\frac{1}{2}&\frac{1}{2}&0\\
0&\frac{1}{2}&\frac{1}{2}\\
\frac{1}{4}&0&\frac{3}{4}\\
\end{pmatrix}}
\end{equation}

Leading to a stationary probability $\left[\frac{1}{4},\frac{1}{4},\frac{1}{2}\right]$; the systems spends half of its time in a state of non-participation and an equal one quarter both as all cooperators or defectors. Intuitively there is a single cycle from full cooperation, which may only be invaded by defectors (under the assumption that $\sigma <\frac{Ncr}{N-1}$). Defectors in turn may only be invaded by non-participants. Once in a state of full non-participation, the population may only \textit{slowly} be invaded by cooperators due to the argument above leading to a fixation probability of $\frac{1}{2}$. Therefore non-participation predominates over long time averages as seen in simulation.

\subsection{Explicit Calculation of Transition Probabilities (Intermediate Imitation Strength)}
\label{hard}

The dynamics of the evolution of cooperation amongst a finite-sized population of agents diverges from the behaviour of mean-field treatments such as replicator dynamics. Now the stochastic effects of mutation become significant~\cite{traulsen}. The fixation probability of an $l$ mutant in an otherwise homogeneous population of $k$ agents, (2), can be calculated explicitly from the theory of birth-death processes~\cite{novakBook} as

\begin{equation}
\rho_{k,l}=\frac{1}{1+\sum_{q=1}^{M-1} \Pi_{N_{l}=1}^{q}\frac{\tau_{l\rightarrow k (N_{l})}}{\tau_{k\rightarrow l (N_{l})}}}
\end{equation}

Where $M$ is the size of the population and the number of agents with strategy $k$ or $l$ respectively is given by $N_{k}$ and $N_{l}$ with $M=N_{k}+N_{l}$. Here $\tau_{l\rightarrow k (N_{l})}$ represents the probability that one of the $N_{k}$ players will convert to strategy $l$ via imitation. This transition probability for a single agent can be written explicitly for a Moran process obeying a logistic imitation probability.

\begin{equation}
	\tau_{l\rightarrow k}(N_{l})=\frac{N_{l}}{M}\frac{M-N_{l}}{M}\frac{1}{1+\exp{[-s(P_{k}-P_{l})]}}
\end{equation}

Where $s$ is the imitation strength and $P_{k}$ and $P_{l}$ are the payoffs of strategies $k$ and $l$ which depend on the number of $k$ and $l$ players. Thankfully the fixation probability simplifies to

\begin{equation}
	\rho_{k,l}=\frac{1}{1+\sum_{q=1}^{M-1} \exp[-s\sum_{N_{l}=1}^{q}(P_{k}-P_{l})]}
\end{equation}

Although there is no analytical expression for this at intermediate values of $s$, the sums can be readily evaluated and the entries of $T$ calculated. In turn the stationary distribution can be calculated.\\ 

Henceforth, unless otherwise specified, we use the following parameter values.
\begin{center}
\begin{tabular}{|c|c|c|}
	\hline
	PGG contribution&c& 1.0\\
	  PGG multiplier&r& 3.0\\
	Population size &M& 100\\
	    Sample size &N& 5\\
      Imitation strength&s&1000\\
	\hline
   Non-participation payoff&$\sigma$& 0.1\\
	    Pool punishment effect&B& 0.7\\
	      Pool punishment cost&G& 0.7\\
	    Peer punishment effect&$\beta$& 0.7\\
	      Peer punishment cost&$\gamma$& 0.7\\
	Bribe as proportion of tax&K&0.5\\
	\hline
\end{tabular}
\end{center}
\clearpage
\section{Replicating Results of Sigmund \textit{et al}}
Sigmund \textit{et al}~\cite{sigmund} calculate the stationary distributions of their simulations in an analagous way. However the introduction of new punishing strategies introduces a fourth possible value for the fixation probability. When a peer-punishing mutant arises in a homogeneous population of cooperators, there is neutral drift since peer-punishers have no-one to punish so enjoy the same benefits as cooperators with no additional costs. This leads to a fixation probability of $\frac{1}{M}$~\cite{novakBook}. In this scheme, the possible strategies are:

\begin{description}
\item[Cooperators ($X$):]  Participate and contribute $c$ to the PGG
\item[Defectors ($Y$):] Participate but do not contribute to the PGG
\item[Loners ($Z$):] Neither participate nor contribute to PGG
\item[Peer Punishers ($W$):] Participate and contribute to the PGG (cooperate) and pay a fixed cost per defector $\gamma$ to punish defectors if encountered (the more the defectors, the more the cost). 
\item[Pool Punishers ($V$):] Participate and contribute to the PGG (cooperate) and pay a fixed a prior cost $G$ toward a punishment pool (central authority), which will punish defectors if defectors appear.
\end{description}

The payoff is determined by choosing a sample population of size $N$ to play the public good game. Below is the payoff calculations for the different strategies. It is important to point out the second order punishment term $B\times V \times \frac{N-1}{M-1}$, where $B$ is a constant that determines the severity of second order punishment.

\begin{eqnarray*}
P_{\sigma} &=& \frac{\binom{Z}{N-1}}{\binom{M-1}{N-1}}\\
P_{second} &=& \frac{\binom{M-Y-2}{N-2}}{\binom{M-2}{N-2}}\\
Y \textrm{ payoff}  &=& (P_{\sigma}.\sigma)  +  (1-P_{\sigma}).r.c.\frac{ M- Z- Y-C}{ M- Z} - B(N-1)\frac{V+H}{M-1}  -  \beta.\frac{(N-1).W+H}{M-1}\\
X \textrm{ payoff}  &=&  (P_{\sigma} \sigma)  +  (1-P_{\sigma}).c.\left(r.\frac{ M- Z- Y-C}{ M- Z}-1\right) - B(N-1)\frac{V+H}{M-1} \\
& & -  \beta.\frac{(N-1).W}{M-1} .(1-P_{second}) \\
Z \textrm{ payoff} &=& \sigma \\
W  \textrm{ payoff}  &=&  (P_{\sigma} \sigma)  +  (1-P_{\sigma}).c. \left(r.\frac{ M- Z- Y-C}{ M- Z}-1 \right) -  (N-1).\frac{Y+C}{M-1}.\gamma  \\
& & - \frac{(N-1)X}{M-1}.\gamma.(1-P_{second} )- B(N-1)\frac{V+H}{M-1}  \\
V \textrm{ payoff}  &=&  (P_{\sigma} \sigma)  +  (1-P_{\sigma}).\left(c. \left[r.\frac{ M- Z- Y-C}{ M- Z}-1 \right] - G \right)\\
\end{eqnarray*}

\noindent The transition matrix is given by:

\begin{equation}
\hspace{-2cm} 
	\bordermatrix{&X&Y&Z&V&W\cr
			X&T_{XX}&T_{XY}&T_{XZ}&T_{XV}&T_{XW}\cr
		     Y&T_{YX}&T_{YY}&T_{YZ}&T_{YV}&T_{YW}\cr
		     Z&T_{ZX}&T_{ZY}&T_{ZZ}&T_{ZV}&T_{ZW}\cr
		     V&T_{VX}&T_{VY}&T_{VZ}&T_{VV}&T_{VW}\cr
		     W&T_{WX}&T_{WY}&T_{WZ}&T_{WV}&T_{WW}\cr
}
\end{equation}

Where

\begin{equation}
	T_{ij}
	\begin{cases}
		\frac{1}{4}\mu\rho_{ij}& \text{if } i\neq j\\
  1-\frac{1}{4}\mu\sum_{k\neq i}\rho_{ik} & \text{if } i=j
	\end{cases}
\end{equation}

\begin{figure}[ht]
\begin{center}
\includegraphics[width=\textwidth]{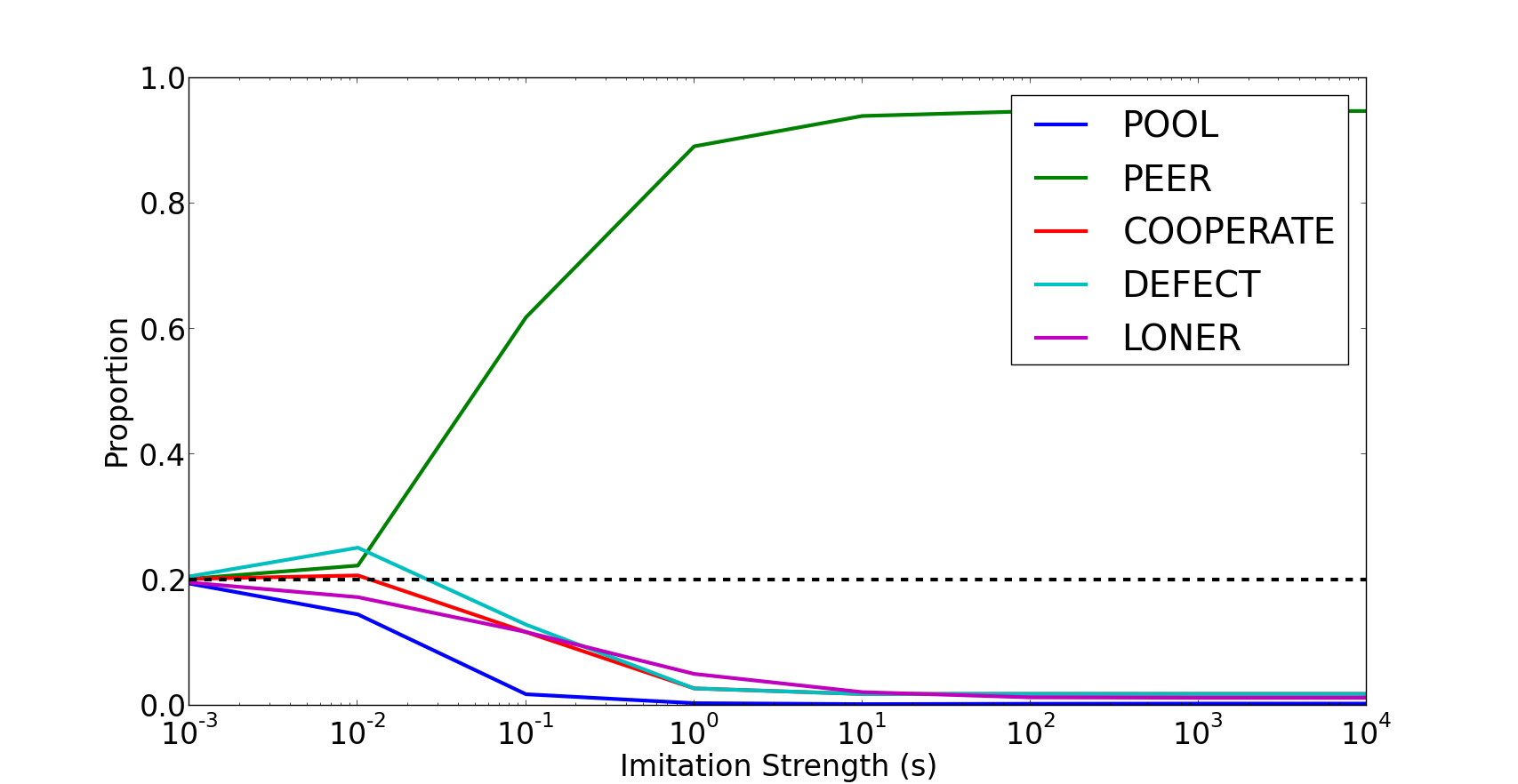}
\caption{Stationary distributions of states as a function of imitation strength. The dashed line represents equal distribution between the $d$ states.}
\end{center}
\label{vwxyz}
\end{figure}

This reduces to

\begin{equation}
	\bordermatrix{&X&Y&Z&V&W\cr
	       X&\frac{3}{4}-\frac{1}{4M}&  \frac{1}{4}&0  & 0 & \frac{1}{4M}\cr
	       Y&0&  \frac{3}{4}&\frac{1}{4}  & 0 & 0\cr
	       Z&\frac{1}{8}& 0&\frac{5}{8}&\frac{1}{8}  & \frac{1}{8}\cr
	       V&\frac{1}{4}& 0&0  & \frac{1}{2} & \frac{1}{4}\cr
	       W&\frac{1}{4M}& 0&0  & 0 & 1-\frac{1}{4M}\cr
}
\end{equation}
With the stationary distribution $\frac{1}{3M+23}\left[6,6,4,1,3M+6\right]$ i.e. Peer-punishers predominate. See Fig(\ref{vwxyz}).\\

Including second order punishment leads to pool punishers dominating. Pool punishers now punish defectors, cooperators and peer punishers for not contributing to the pool. Peer-punishers continue to punish defectors and cooperators.\\

The main differences introduced is that there is no longer a neutral drift between cooperators and peer punishers ($\rho_{X,W}\rightarrow 0$), cooperators no longer invade pool-punishers ($\rho_{V,X}\rightarrow 0$) or peer-punishers ($\rho_{V,W}\rightarrow 0$).\\

The transition matrix becomes

\begin{equation}
	\bordermatrix{
		&X&Y&Z&V&W\cr
	       X&\frac{3}{4}&  \frac{1}{4}&0  & 0 &0\cr
	       Y&0&  \frac{3}{4}&\frac{1}{4}  & 0 & 0\cr
	       Z&\frac{1}{8}& 0&\frac{5}{8}&\frac{1}{8}  & \frac{1}{8}\cr
	       V&0& 0&0  & 1 & 0\cr
	       W&\frac{1}{4M}& 0&0  & 0 & 1-\frac{1}{4M}\cr
}
\end{equation}

\begin{figure}[ht]
\begin{center}
\includegraphics[width=\textwidth]{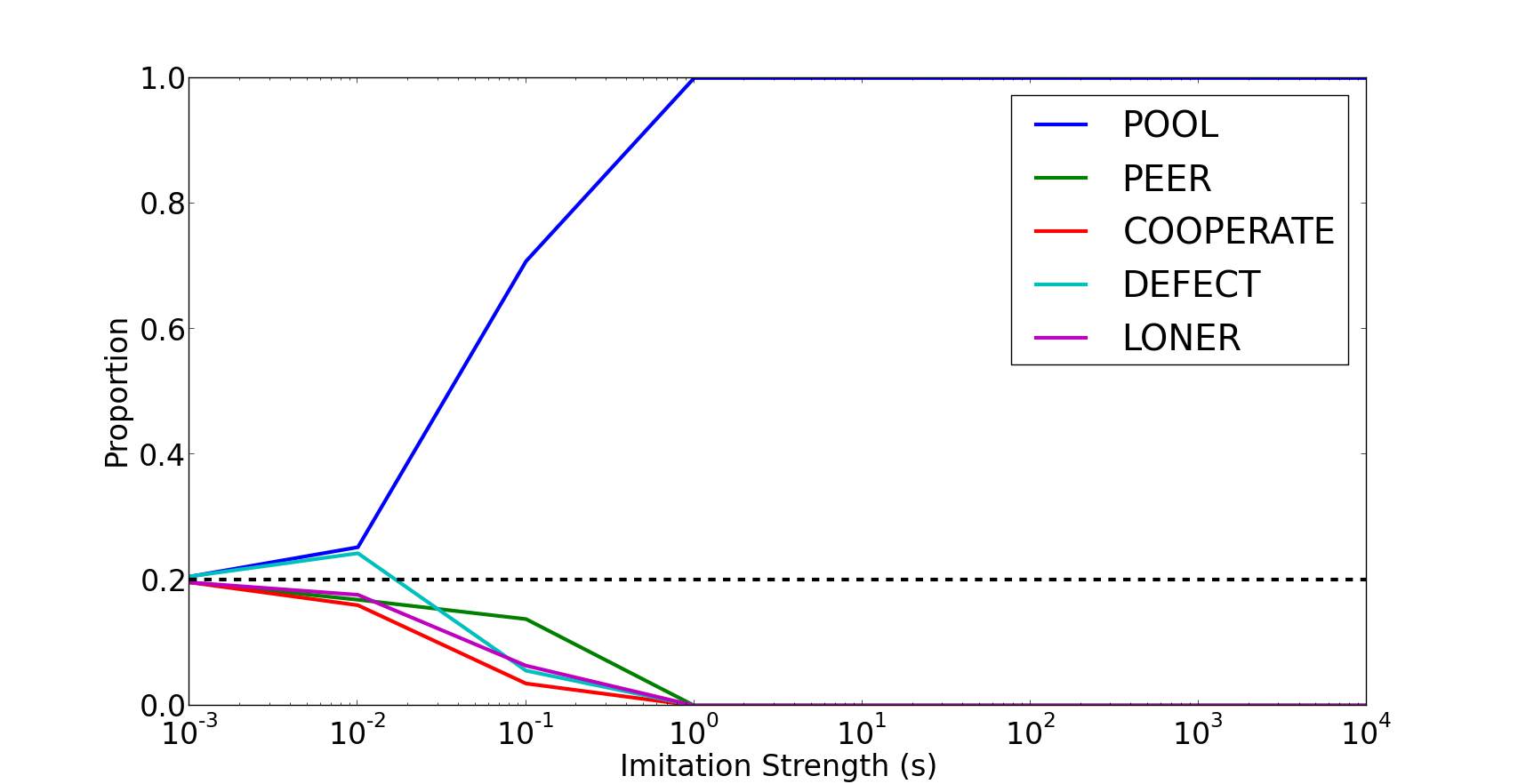}
\caption{Stationary distributions of states as a function of imitation strength. The dashed line represents equal distribution between the $d$ states.}
\end{center}
\label{vwxyz_second}
\end{figure}

Since there is no flow out of a state of full pool-punishers, but flow into it; the stationary distribution becomes $\left[ 0,0,0,1,0\right]$. (See Fig(\ref{vwxyz_second})). Thus the presence of second-order punishment of second-order free-riders (cooperators and peer-punishers) determines whether pool-punishers or peer-punishers will prevail. The latter outcome is preferable since pool-punishers have clear dominance, whereas without second order punishment cooperation is susceptible to breaking down (See~\cite{sigmund} Fig 3a, main paper)

\clearpage

\section{Corruptors}
\label{corruptors}
We now introduce a fifth strategy into the model of Sigmund \textit{et al}:

\begin{description}
\item[Corruptors ($C$):] A corruptor pays the central authority a fixed fee $KG<G+c$ to avoid punishment for defecting from the PGG. Parameter $K\in[0,1]$ here is a new parameter that controls bribe as percentage of $G$, the fee paid by pool punishers (well-behaving citizens). 
\end{description}

\noindent The payoff of corruptors is:

\begin{eqnarray*}
C \textrm{ payoff}  &=&  (P_{\sigma} \sigma)  +  (1-P_{\sigma}).\left(c. r.\frac{ M- Z- Y-C}{ M- Z}  - KG \right) -  \beta.(N-1)\frac{W+H}{M-1}
\end{eqnarray*}

This leads to the larger transition matrix:

\begin{equation}
\hspace{-2cm} 
\bordermatrix{&X&Y&Z&V&W&C\cr
		 X&T_{XX}&T_{XY}&T_{XZ}&T_{XV}&T_{XW}&T_{XC}\cr
		    Y&T_{YX}&T_{YY}&T_{YZ}&T_{YV}&T_{YW}&T_{YC}\cr
		    Z&T_{ZX}&T_{ZY}&T_{ZZ}&T_{ZV}&T_{ZW}&T_{ZC}\cr
		    V&T_{VX}&T_{VY}&T_{VZ}&T_{VV}&T_{VW}&T_{VC}\cr
		    W&T_{WX}&T_{WY}&T_{WZ}&T_{WV}&T_{WW}&T_{WC}\cr
		    C&T_{CX}&T_{CY}&T_{CZ}&T_{CV}&T_{CW}&T_{CC}\cr
}
\end{equation}

\subsection{Weak Pool Punishment (Low $B$)}
When second-order punishment is weak (low values of $B$), peer punishers are stable with respect to pool-punishers. Substitution for the fixation probabilities leads to 

\begin{equation}
	\label{corruptMatrixLow}
	\bordermatrix{
		&X&Y&Z&V&W&C\cr
	X&\frac{3}{5}&  \frac{1}{5}&0  & 0 &0&\frac{1}{5}\cr
	Y&0&  \frac{4}{5}&\frac{1}{5}  & 0 & 0&0\cr
	       Z&\frac{1}{10}& 0&\frac{7}{10}&\frac{1}{10}  & \frac{1}{10}&0\cr
	       V&0& 0&0  & \frac{4}{5} & 0&\frac{1}{5}\cr
	       W&\frac{1}{5M}& 0&0  & 0 & 1-\frac{1}{5M}\cr
	       C&0&\frac{1}{5}& \frac{1}{5}&0  & 0 & \frac{3}{5}\cr
}
\end{equation}

The stationary distribution is now $\frac{1}{M+7}\left[1,2,2,1,M,1\right]$ (using a population size $M=100$ this is approximately $\left[ 0.01,0.02,0.02,0.01,0.93,0.01\right]$) confirming clear dominance of peer-punishers. 
\subsection{Strong Pool Punishment (High $B$)}
However, under extremely high second-order punishment cooperation breaks down with pool punishers dominating followed by loners and corrupt. Modifying (\ref{corruptMatrixLow}) yields

\begin{equation}
	\label{corruptMatrixHigh}
	\bordermatrix{
		&X&Y&Z&V&W&C\cr
	X&\frac{2}{5}-\frac{1}{5M}&  \frac{1}{5}&0  & \frac{1}{5} &\frac{1}{5M}&\frac{1}{5}\cr
	Y&0&  \frac{3}{5}&\frac{1}{5}  & \frac{1}{5} & 0&0\cr
	       Z&\frac{1}{10}& 0&\frac{7}{10}&\frac{1}{10}  & \frac{1}{10}&0\cr
	       V&0& 0&0  & \frac{4}{5} & 0&\frac{1}{5}\cr
	       W&\frac{1}{5M}& 0&0  & \frac{1}{5} & \frac{4}{5}-\frac{1}{5M}&0\cr
	       C&0&\frac{1}{5}& \frac{1}{5}&0  & 0 & \frac{3}{5}\cr
}
\end{equation}

This leads to a stationary distribution of 

\begin{multline}
	\hspace{-1.1cm}
\frac{1}{\frac{77}{16}+\frac{33(2+3M)}{16(22+17M)}} \Biggl[ \frac{3}{8}-\frac{9(2+3M)}{8(22+17M)},\frac{11}{16}-\frac{9(2+3M)}{16(22+17M)}, \\ \frac{9}{8}-\frac{3(2+3M)}{8(22+17M)}, \frac{13}{8}+\frac{9(2+3M)}{8(22+17M)},\frac{3(2+3M)}{(22+17M)},1 \Biggr]
\end{multline}

This can be evaulated with $M=100$ as $\left[0.034,0.114,0.204,0.352,0.102,0.193\right]$ i.e. pool-punishers predominate, followed by loners and corruptors.

\begin{figure}[ht]
\begin{center}
\includegraphics[width=\textwidth]{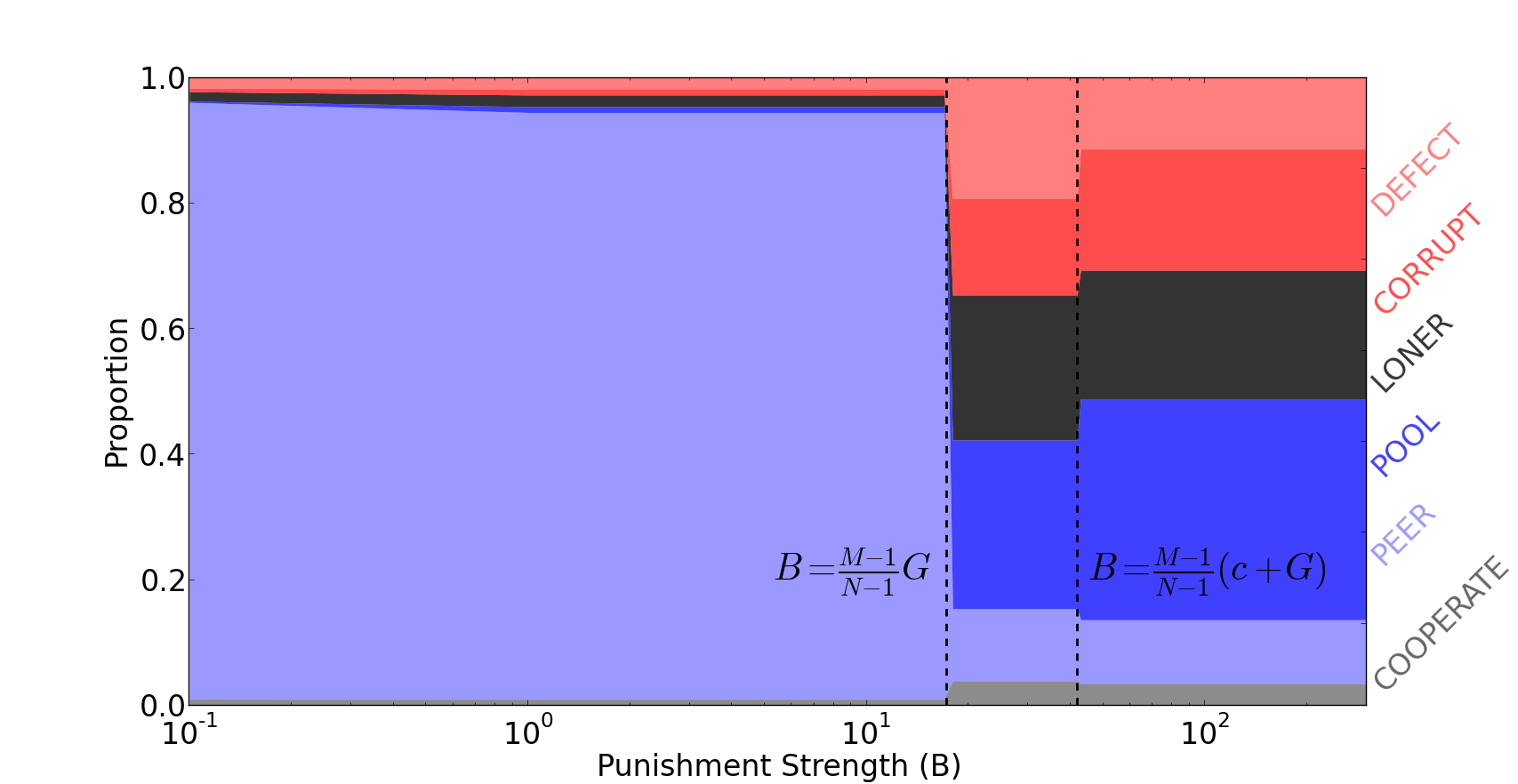}
\caption{Stationary distributions of states as a function of pool punishment strength. As $B$ increases cooperation breaks down.}
\end{center}
\label{fig:xyzvwch}
\end{figure}

We see 2 very clear discontinuities at $B\approx17$ and $B\approx40$ when the proportion of peer punishers drops to be replaced by pool-punishers. Above the first threshold, pool-punishing agents will invade peer-punishers ($\rho_{WV}\rightarrow1$). Above the second threshold, pool-punishing agents will also take over defectors. These points are explained below.\\

Firstly, at intermediate values of $B$ the expected pool-punishment (calculated from the probability of being selected with the single pool-punisher in the sample of $N$) is low. Therefore, the low probability of being matched with a pool-punisher doesn't incentivise the payment of $G$. However, once $B$ is sufficiently high, the threat of pool-punishment \textit{even from a single pool-punishing hybrid player} is too high of a risk and all non-pool-punishing strategies can be invaded by pool-punishing strategies ($\rho_{WH},\rho_{WV}\rightarrow 1$). The condition for this is given by the expected cost of receiving pool-punishment when a single pool-punishing hybrid agent is present in a population

\begin{equation}
	(\frac{N-1}{M-1})B
\end{equation}

When this is equal to $G$ it is cheaper to pay tax than to risk pool-punishment

\begin{equation}
	G<\frac{(N-1)}{M-1}B
\end{equation}

\begin{equation}
	B^{*}=(\frac{M-1}{N-1})G
\end{equation}

Substituting $N=5$,$M=100$ and $G=0.7$ gives a critical value when $B^{*}=17.325$.\\

Adressing the second threshold; as the pool-punishment term becomes very large, the expected value of pool-punishment for a homogeneous population of defectors being punished by a single pool-punisher becomes so large that pool-punishers may invade defectors, despite the pool-punisher making a heavy loss in the PGG.

\begin{equation}
	c+G<\frac{N-1}{M-1}B
\end{equation}

\begin{equation}
	B^{*}=\frac{M-1}{N-1}(c+G)
\end{equation}

Substituting for $M,N,c$ and $G$ gives $B^{*}=42.075$.

\clearpage

\section{Corruptors and Hybrid Punishers}
Finally, we add Hybrid-Punishers (H) to the set of possible strategies. 

\begin{description}
\item[Hybrid-Punishers ($H$):] These players participate and contribute to the PGG (cooperate), pay a fixed cost per defector $\gamma$, and pay a fixed a prior cost $G$ toward a punishment pool. 
\end{description}

\noindent The payoff of hybrid punishers is then defined by the following equation:

\begin{eqnarray*}
H \textrm{ payoff}  &=&  (P_{\sigma} \sigma)  +  (1-P_{\sigma}).\left(c. \left[ r.\frac{ M- Z- Y-C }{ M- Z}-1 \right]  - G \right) -  (N-1).\frac{Y+C}{M-1}.\gamma
\end{eqnarray*}

\noindent We now have the following transition matrix.

\begin{equation}
\hspace{-2cm} 
\bordermatrix{&X&Y&Z&V&W&C&H\cr
				  X&T_{XX}&T_{XY}&T_{XZ}&T_{XV}&T_{XW}&T_{XC}&T_{XH}\cr
		    Y&T_{YX}&T_{YY}&T_{YZ}&T_{YV}&T_{YW}&T_{YC}&T_{YH}\cr
		    Z&T_{ZX}&T_{ZY}&T_{ZZ}&T_{ZV}&T_{ZW}&T_{ZC}&T_{ZH}\cr
		    V&T_{VX}&T_{VY}&T_{VZ}&T_{VV}&T_{VW}&T_{VC}&T_{VH}\cr
		    W&T_{WX}&T_{WY}&T_{WZ}&T_{WV}&T_{WW}&T_{WC}&T_{WH}\cr
		    C&T_{CX}&T_{CY}&T_{CZ}&T_{CV}&T_{CW}&T_{CC}&T_{CH}\cr
		    H&T_{HX}&T_{HY}&T_{HZ}&T_{HV}&T_{HW}&T_{HC}&T_{HH}\cr
}
\end{equation}

\subsection{Weak Pool Punishment (Low $B$)}

Assuming a low value of $B$, results in the transition matrix below.

\begin{equation}
\label{corruptHybridMatrixLow}
\bordermatrix{
&X&Y&Z&V&W&C&H\cr
X&\frac{3}{6}-\frac{1}{6M}&\frac{1}{6}&0&\frac{1}{6}&\frac{1}{6M}&\frac{1}{6}&0\cr
Y&0&\frac{4}{6}&\frac{1}{6}&\frac{1}{6}&0&0&0\cr
Z&\frac{1}{12}&0&\frac{9}{12}&\frac{1}{12}&\frac{1}{12}&0&0\cr
V&0&0&0&\frac{5}{6}-\frac{1}{6M}&0&\frac{1}{6}&\frac{1}{6M}\cr
W&\frac{1}{6M}&0&0&0&1-\frac{1}{6M}&0&0\cr
C&0&\frac{1}{6}&\frac{1}{6}&0&0&\frac{2}{3}&0\cr
H&\frac{1}{6}&\frac{1}{6}&\frac{1}{6}&\frac{1}{6M}&\frac{1}{6}&\frac{1}{6}&\frac{1}{6}-\frac{1}{6M}\cr
}
\end{equation}

The stationary distribution in this case is given as 

\begin{equation}
	\frac{1}{\Gamma}\left[ \frac{24}{13}+\frac{15M}{13},\frac{31}{15}+\frac{55M}{26},\frac{46}{13}+\frac{45M}{13},1+5M,3(16+34M+15M^{2},5(5+8M)),1  \right]
\end{equation}

Where the normalisation factor is given as 

\begin{equation}
	\Gamma=\frac{127}{13}+\frac{305M}{26}+\frac{5}{13}(5+8M)+\frac{3}{26}(16+34M+15M^{2})
\end{equation}

This evaluates to $\left[0.01,0.017,0.016,0.008,0.94,0.006,0.001 \right]$. Peer punishers overwhelmingly predominate, followed by defectors, loners and cooperators (agrees with low $B$ limit of Fig 4 of corruption paper).\\
\subsection{Strong Pool Punishment (High $B$)}

When $B$ is very large the transition matrix becomes

\begin{equation}
\label{corruptHybridMatrixHigh}
\bordermatrix{
&X&Y&Z&V&W&C&H\cr
X&\frac{2}{6}-\frac{1}{6M}&\frac{1}{6}&0&\frac{1}{6}&\frac{1}{6M}&\frac{1}{6}&\frac{1}{6}\cr
Y&0&\frac{4}{6}&\frac{1}{6}&\frac{1}{6}&0&0&0\cr
Z&\frac{1}{12}&0&\frac{2}{3}&\frac{1}{12}&\frac{1}{12}&0&\frac{1}{12}\cr
V&0&0&0&\frac{5}{6}-\frac{1}{6M}&0&\frac{1}{6}&\frac{1}{6M}\cr
W&\frac{1}{6M}&0&0&0&\frac{5}{6}-\frac{1}{6M}&0&\frac{1}{6}\cr
C&0&\frac{1}{6}&\frac{1}{6}&0&0&\frac{2}{3}&0\cr
H&0&0&0&\frac{1}{6M}&0&0&1-\frac{1}{6M}\cr
}
\end{equation}

The stationary distribution can be expressed as

\begin{multline}
\frac{1}{\Gamma} \Biggl[ \frac{3(2+M)}{(70+86M+27M^{2})}, \frac{-22-17M}{(70+86M+27M^{2})}, \frac{6(5+4M)}{(70+86M+27M^{2})}, \\ \frac{-70-59M}{(70+86M+27M^{2})}, \frac{6(1+2M)}{(70+86M+27M^{2})}, \frac{-38-31M}{(70+86M+27M^{2})}, 1 \Biggr]
\end{multline}

Where the normalisation factor is given as 

\begin{multline}
\Gamma=\frac{1}{\splitfrac{1-\frac{-70-59M}{70+86M+27M^{2}} -\frac{-38-31M}{70+86M+27M^{2}} -\frac{-22-17M}{70+86M+27M^{2}}-\frac{3(2+M)}{70+86M+27M^{2}}-\frac{6(1+2M)}{70+86M+27M^{2}}}{-\frac{6(5+4M)}{70+86M+27M^{2}} } }
\end{multline}

With the stationary distribution as follows $\left[ 0.001, 0.0059, 0.008, 0.020, 0.004, 0.011, 0.950\right]$; hybrid punishers predominate (in agreement with the high $B$ limit of Fig 4 of corruption paper). The proportions are plotted as a function of $B$ below in Fig (\ref{fig:xyzvwch}). \\

\begin{figure}[ht]
\begin{center}
\includegraphics[width=\textwidth]{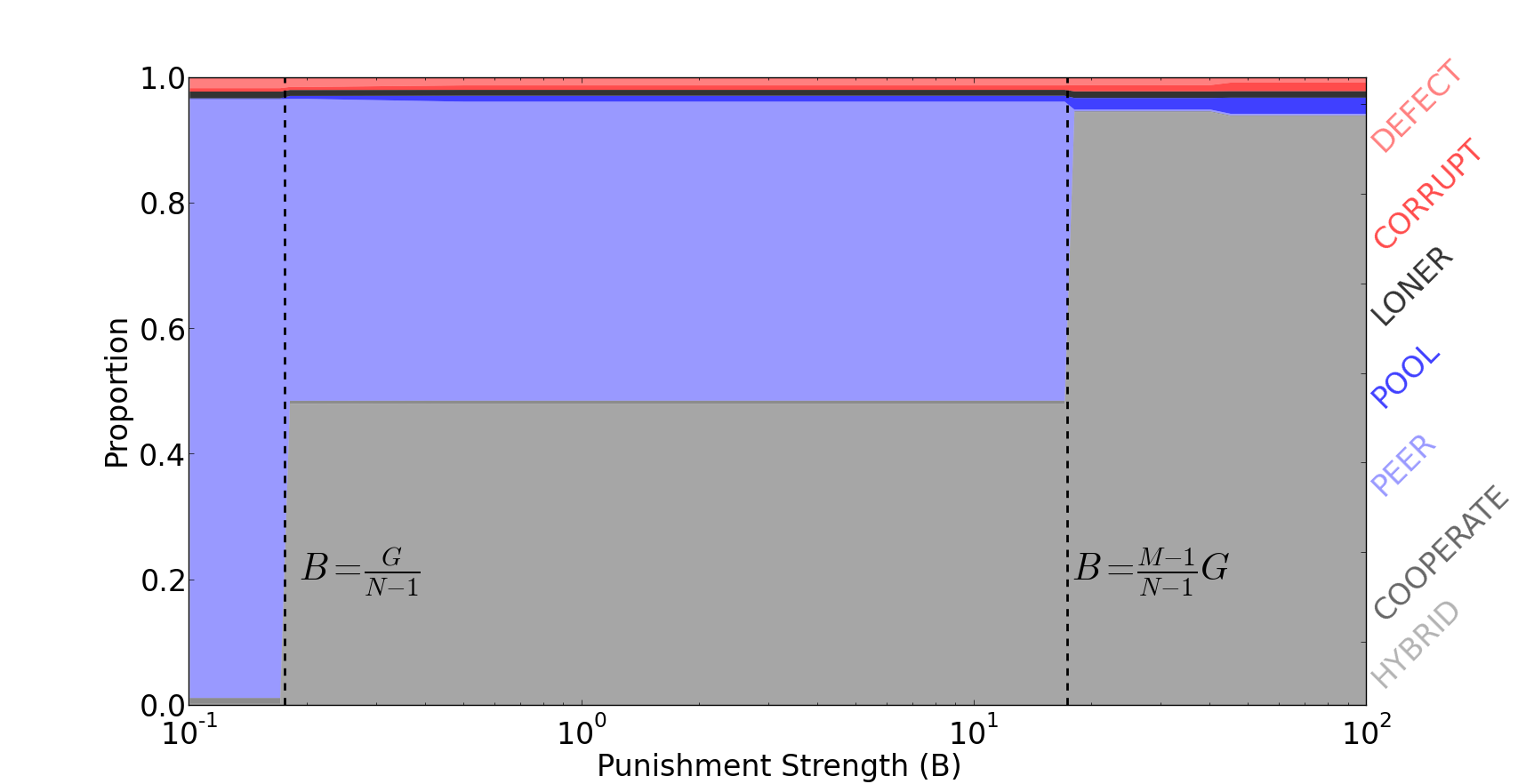}
\caption{Stationary distributions of states as a function of pool punishment strength. As $B$ increases hybrid punishers become dominant.}
\end{center}
\label{fig:xyzvwch}
\end{figure}

We see 2 very clear discontinuities at $B\approx0.2$ and $B\approx17$ when the proportion of peer punishers drops to be replaced by hybrid strategies. Above the first threshold; hybrid strategies may no longer be invaded by peer-punishers ($\rho_{HW}\rightarrow0$). Above the second threshold, hybrid agents will also invade peer-punishers ($\rho_{WH}\rightarrow1$). The explanation for the second threshold is the same as section \ref{corruptors} and the first threshold is explained below.\\

For a single peer-punishing mutant to invade hybrid players, the saving from paying the tax $G$ must outweigh any possible second order pool-punishment. Since, apart from the mutant herself, only pool-punishers are present this has an expected value of $B(N-1)$ i.e. punishment from all the other players in the sample.

\begin{equation}
G<B(N-1)
\end{equation}

Leading to a threshold value for $B^{*}=0.175$.\\

The transition matrix in (\ref{corruptHybridMatrixHigh}) also shows that when second-order punishment is strong, hybrid punishers are only destablized by neutral drift towards pool-punishers, who can then be exploited by corruptors. One interpretation is that this form of instability represents a risk that exists in the real world. When there is high cooperation, individuals might become lax in their propensity to altruistically punish defection, and this can destablize cooperation. As mentioned in the main text, this risk may motivate goverments to sometimes mandate that citizens to sign up for certain peer punishment duties, like jury duty, and punish those who merely pay their taxes. If pool-punishers were also punished by second-order punishment, then there would be no neutral drift towards this strategy, and the stationary distribution would be  $\left[ 0,0, 0, 0, 0, 0, 1\right]$, as there would be no flows away from the hybrid punisher state.
